\documentclass[aps,prl,twocolumn,superscriptaddress,nofootinbib]{revtex4-1}
\usepackage{amssymb, graphicx, hyperref, subcaption}
\usepackage[intlimits]{amsmath}
\usepackage[english]{babel}

\begin{document}

\title{Dissociation of heavy quarkonium states in rapidly varying strong magnetic field}
\author{Partha Bagchi}
\email[E-mail:]{p.bagchi@vecc.gov.in}
\affiliation{Theoretical Physics Division, Variable Energy Cyclotron Centre,
1/AF, Bidhan Nagar, 
Kolkata 700064, India}

\author{Nirupam Dutta}
\email[E-mail:]{nirupamdu@gmail.com}
\affiliation{School of Physical Sciences, National Institute of Science Education and Research Bhubaneswar,
P.O. Jatni, Khurda 752050, Odisha, India}

\author{Bhaswar Chatterjee}
\email[E-mail:]{bhaswar.mph2016@iitr.ac.in}
\affiliation{Department of Physics, Indian Institute of Technology Roorkee, Roorkee 247667, India}

\author{Souvik Priyam Adhya}
\email[E-mail:]{sp.adhya@vecc.gov.in}
\affiliation{Theoretical Physics Division, Variable Energy Cyclotron Centre,
1/AF, Bidhan Nagar, 
Kolkata 700064, India}

\date{\today}

\begin{abstract}
In a transient magnetic field, heavy quarkonium bound states evolve non adiabatically. In presence of a strong magnetic field,  $J/\Psi$ and $\Upsilon(1S)$ become more tightly bound than we expected earlier for a pure thermal medium. We have shown that in a time varying magnetic field, there is a possibility of moderate suppression of  $J/\Psi$ through the non adiabatic transition to continuum where as the $\Upsilon(1S)$ is so tightly bound that can not be dissociated through this process. We have calculated the dissociation probabilities up to the first order in the time dependent perturbation theory for different values of initial magnetic field intensity.
\end{abstract}

\maketitle

In recent time, it has been argued that a very high intensity magnetic field is expected \cite{skokov2009,kharzeev2007,Mueller:2014tea,Miransky:2015ava} to be formed in non central high energy nucleus-nucleus collisions. This realisation already has motivated several investigations searching interesting perturbative and non-perturbative phenomena \cite{Kharzeev:2007tn,bloczynski2012,Asakawa:2010bu} of QCD matter in the laboratory. On the other hand, the magnetic field can modify several issues dramatically which previously have been understood without it. For example, the issue of heavy quarkonia suppression in the deconfined Quark Gluon Plasma \cite{bonati2015,guo2015,yang2011,} can be modified greatly if one considers the magnetic field into account. A very obvious modification in this area is the Zeeman splitting of quarkonium states in constant magnetic field which essentially creates various quarkonium states \cite{filip2013,michael2013} differing by their spin degrees of freedom which is very similar to the case of positronium in quantum electrodynamics \cite{karshenboim2003}. Then, there are possibilities for spin mixing in homogeneous \cite{michael2013,yang2011} and inhomogeneous \cite{Dutta:2017pya} magnetic field environment. Besides that, ionisation \cite{tuchin2011} of bound states due to the tunnelling caused by the magnetic field can lead to suppression of quarkonium states. Furthermore, the static quark anti-quark potential in medium can be modified up to a big extant if the magnetic field can persist for a longer time. Depending on non-centrality, the magnetic field can be as strong as $B \simeq$ 50 $m_\pi^2$ where $m_\pi^2=10^{18}$ Gauss. This field strength decays very quickly as the spectator quarks move away from the fireball and it has been estimated that at time $t \simeq$ 0.4 $fm$, the magnetic field is practically negligible. However, if QGP is formed, then it can trap the magnetic field because of its high electrical conductivity. So the formation of QGP can increase the persistence time \cite{ajit2017} of magnetic field in Relativistic Heavy Ion Collision (RHIC). Nevertheless, the field will decay to few orders of magnitude within few $fm/c$ time. Hence, the produced magnetic field is time dependent and in turn would significantly affect the production of particles and their subsequent dynamics. So it is worth studying the properties of quarkonia in presence of such transient (or time varying) magnetic field.

This is true that there are several view points regarding the nature of the magnetic field generated through Heavy Ion Collisions (HIC) and hence, whatever we predict at the moment by considering the speculative ideas of the magnetic field may not lead us to a proper quantitative predictions of observables. Nevertheless, the qualitative aspects of various phenomena can be understood well enough. In this article, we have considered a magnetic field which is decaying with time and have calculated the transition of quarkonia to the continuum states from the bound one. This leads to further suppression of quarkonia which is completely different from the ionisation process discussed earlier \cite{tuchin2011}. In a time varying magnetic field, quarkonia evolves non-adiabatically because the quark anti-quark potential becomes time dependent and changes very rapidly as the magnetic field does. We have investigated the time evolution of spatial wave functions of quarkonia and therefore have not considered the spin-magnetic field interaction into account for the current article. The non-adiabatic evolution previously has been addressed in the context of evolving QGP \cite{Dutta:2012nw} and also in the context of rapid thermalisation \cite{Bagchi:2014jya}.

 In this work, we will restrict ourselves within the strong magnetic field approximation which essentially means that the magnetic field will act as the dominant 
scale and will prevail over other scales present in the system such as mass and temperature as because $\frac{eB}{m^2} >> 1$ and $\frac{eB}{T^2}>>1$, where $m$ is the mass of the particle affected by magnetic field and $T$ is the temperature of the system. This is obviously above the Schwinger's critical limit \cite{Schwinger1951} that makes it possible to have a classical description of the magnetic field. The effects of magnetic field is incorporated through the propagator of the charged particles present in the medium which in our case are the light quarks. Though there is no effect of magnetic field on the gluon propagator at the zeroth order, it gets affected in the next order through vacuum fluctuation. The fermion propagator in the strong field limit is given by 
\begin{equation}
S_0(k)= i\frac{m + \gamma\cdot k_\parallel}{k_\parallel^2-m^2}(1-i\gamma_1\gamma_2)
e^{\frac{-k_\perp^2}{|q_fB|}}
\end{equation}

for zero temperature. Here we have assumed the magnetic field, B to be along a fixed direction (lets say $z$). 
$q_f$ is he electric charge of the fermion of flavor $f$ and $K$ is the fermion 
4-momentum expressed as $k_\perp^2= -(k_x^2+k_y^2)$, $k_\parallel^2 = k_0^2 + k_z^2$ 
and $\gamma\cdot k_\parallel = \gamma_0 k_0 - \gamma_3 k_z$. The split in the 
4-momentum occurs due to the Landau quantization in the plane transverse to 
the magnetic field as the fermion energy is given by

\begin{equation}
E = \sqrt{m^2 + k_z^2 + 2n|q_f|B}
\end{equation}
with $n$ being the number of Landau levels which is equal to zero in the 
strong field limit. At finite temperature, the propagator in real time \cite{Hasan:2017fmf} becomes

\begin{align}
& iS_{11}(p)=\Bigg[\frac{1}{{p_{\parallel}^2-m^2+
i\epsilon}}+2\pi in_{p}\delta(p_{\parallel}^2-m^2)\Bigg]
(1+\gamma^{0}\gamma^{3}\gamma^{5})\nonumber \\
&\times (\gamma^{0}p_{0}-\gamma^{3}
p_{z}+m) e^{\frac{-p_{\perp}^2}{\mid qB \mid}},
\end{align}

where the distribution is 

\begin{equation*}
n_p(p_0) = \frac{1}{e^{\beta\mid p_0\mid} + 1},
\end{equation*}

with the Bolthzman factor $\beta$. The Debye screening mass ($m_D$) heavy quark potential  in strong magnetic field can be obtained by taking the static limit $(|\vec{p}|=0, p_0 \rightarrow 0)$ of the longitudinal part of the gluon self energy $\pi_{m_D\nu}$. If there is no magnetic field in medium then $m_D$ can be written for three flavor case as $m_D = gT \sqrt{1 + N_f/6} $ \cite{karsch}. In presence of magnetic field, The Debye mass \cite{Hasan:2018kvx} becomes, 

\begin{equation}
m_D^2 = {g^\prime}^2 T^2 + 
\frac{g^2}{4\pi^2T}\sum_f \mid q_fB\mid 
\int_0^\infty dp_z \frac{e^{\beta\sqrt{p_z^2 + m_f^2}}}
{\left(1+e^{\beta\sqrt{p_z^2 + m_f^2}}\right)^2}\label{Debyemass}
\end{equation}

Where the first term is the contribution from the gluon loops and this 
is solely dependent on temperature and magnetic field doesn't affect it. The second term is the contribution from the fermion loop and this term strongly depends on magnetic field and is not much sensitive to the temperature of the medium. In the first term, ${g^\prime}^2 = 4\pi\alpha_s^\prime(T)$ 
where $\alpha_s^\prime(T)$ is the usual temperature dependent running 
coupling where the renormalization scale is taken as $2\pi T$. It is 
given by 

\begin{equation}
\alpha_s^\prime(T) = \frac{2\pi}{\left(11-
\frac{2}{3}N_f\right)\ln\left(\frac{\Lambda}{\Lambda_{QCD}}\right)}
\end{equation}

Where $\Lambda = 2\pi T$ and $\Lambda_{QCD} \sim 200$ MeV

In the second term, $g^2 = 4\pi\alpha_s^\parallel (k_z, q_fB)$, where 
$\alpha_s^\parallel (k_z, q_fB)$ is the magnetic field dependent 
coupling and doesn't depend on temperature. This is given by \cite{Andreichikov:2012xe,Ferrer:2014qka}

\begin{equation}
\alpha_s^\parallel (k_z, q_fB) = \frac{1}
{{\alpha_s^0(\mu_0)}^{-1} + \frac{11 N_c}{12\pi}
\ln\left(\frac{k_z^2 + M_B^2}{\mu_0^2}\right) + \frac{1}{3\pi}
\sum_f \frac{q_fB}{\sigma}}
\end{equation}

where 

\begin{equation}
\alpha_s^0(\mu_0) = \frac{12\pi}{11 N_c \ln \left(\frac{\mu_0^2 + M_B^2}
{\Lambda_V^2}\right)}
\end{equation}

All the parameters are taken as $M_B$ = 1 GeV, the string tension $\sigma$ 
= 0.18 ${GeV}^2$, $\mu_0$ = 1.1 GeV and $\Lambda_V$ = 0.385 GeV.

In the strong field limit, the temperature dependence of the Debye mass is almost negligible. Now one has to see the nature of the magnetic field which decreases with time and that essentially makes the Debye screening mass a time dependent quantity. The intensity of the initial magnetic field $B_0$ is of the order of few $m_{\pi}^2$ and decays with time in the following way, 
\begin{equation}
 B = B_0 \frac{1}{1+a t},
 \label{Btrend}
\end{equation}
using the fitting of the result provided in the article by K. Tuchin \cite{Tuchin:2015oka} with the value of the parameter $a=0.5$. 

The heavy quark potential in medium can be written as,
\begin{equation}
V(r) =  - \frac{\alpha}{r} exp(-m_D r) +
        \frac{\sigma}{ m_D} (1 - exp(-m_D r)) 
\end{equation}
The effect of the temperature and magnetic field is incorporated in the Debye mass given in eq.\ref{Debyemass} This is obvious that potential becomes time dependent due to the time dependence of the magnetic field and temperature. We consider that initially at $t=t_i$, there are only ground states of charmonia ($J/\Psi$) and bottomonia ($\Upsilon(1S)$). These two states evolve in a time dependent potential which causes transition to other excited states and as well as to the dissociated continuum. We would like to calculate the transition probabilities of the ground states to the continuum which gives us the dissociation probabilities of ($J/\Psi$) and ($\Upsilon(1S)$). This is a very difficult task because solving Schr\"{o}edinger equation for a time dependent potential is cumbersome. We have adopted time dependent perturbation theory in this context in order to calculate the dissociation probability up to the first order. The perturbation at any instant $t$ considered to be as $H^1(t)=V(r,t)-V(r,t_i)$. We want to calculate the transition probability to the unbound states which are obviously plane wave states given by,
\begin{equation}
\Psi_{k} =\frac{1}{\sqrt{\Omega}}e^{i\vec{k}.\vec{r}}
\end{equation}
which is box normalised over a volume $\Omega$ and can have all possible values of the momentum $\vec{k}$. The first order contribution to the transition amplitude can be expressed as,
\begin{equation}
 a_{ik}=\int\frac{d}{dt}\langle \Psi_k|H^1(t)|\Psi_i\rangle\frac{e^{i(E_i-E_k)}}{(E_i-E_k)} dt .
\end{equation}
$|\Psi_i\rangle$, $E_i$ are initial quarkonium state and the corresponding energy eigenstates respectively and $E_k$ is the energy of the dissociated state $|\Psi_k\rangle$. The total transition probability to all continuum states is given by,
\begin{equation}
 =\int_{k=0}^{\infty}|a_{ik}|^2\frac{\Omega}{(2\pi)^3} k^2 dk ,
\end{equation}
where the number of unbound states between the momentum continuum $k$ and $k+dk$ over  $4\pi$ solid angle is
\begin{equation}
 dn=\left(\frac{L}{2\pi}\right)^3 k^2 dk =\frac{\Omega}{(2\pi)^3} k^2 dk 
\end{equation}

\begin{figure}
\includegraphics[width=0.5\textwidth]{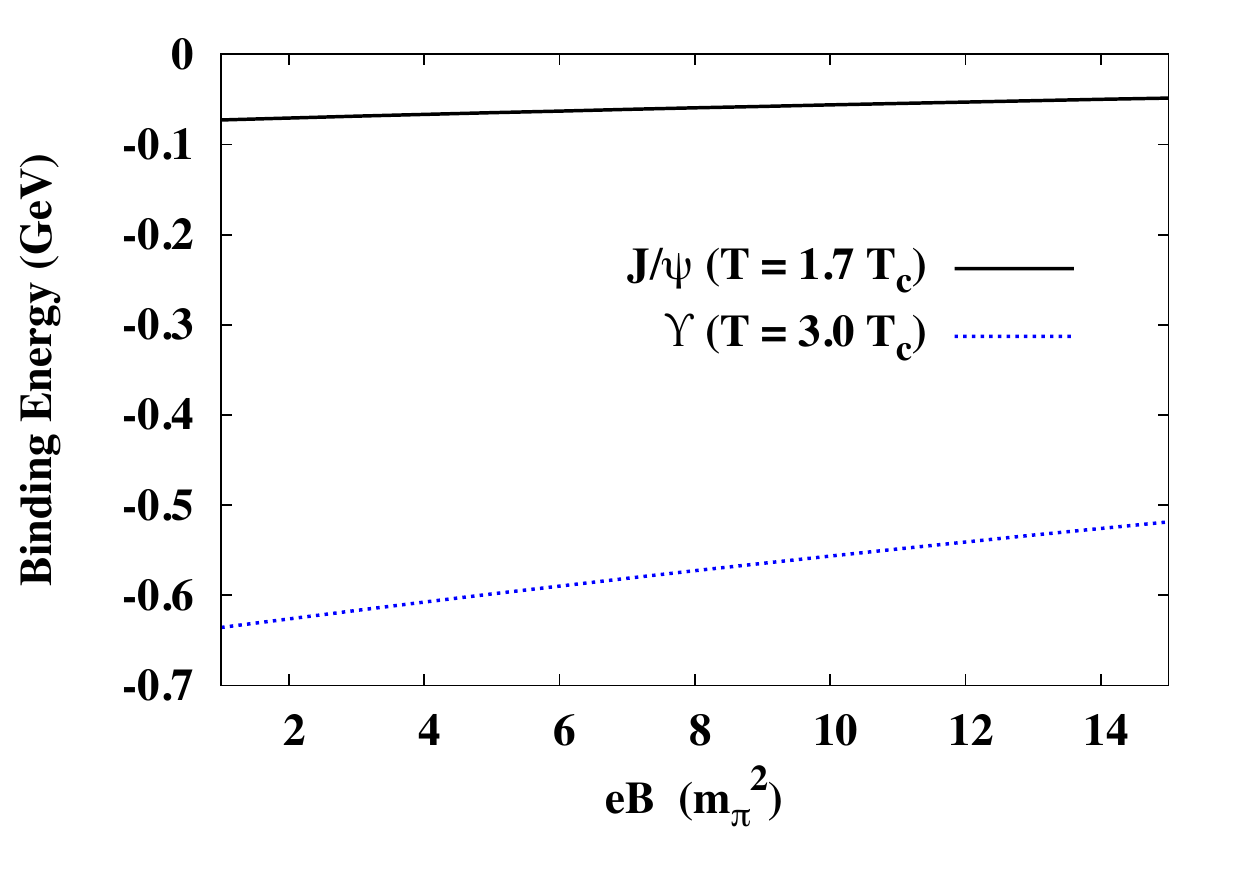}
\caption{Binding energy of $J/\Psi$ and $\Upsilon(1S)$ as a function of the magnetic field intensity.}
\label{binding}
\end{figure}

 We know that $J/\Psi$ and $\Upsilon(1S )$can survive in the thermal medium (QGP) almost upto $2.2 T_c$ and $4T_c$ respectively \cite{PhysRevLett.99.211602} but in the presence of magnetic field, the binding energies of these states get modified. The binding energy is given by,
 \begin{equation}
E_{disso}=E_B-2 m_q-\frac{\sigma}{m_D} ,
\end{equation}\label{eq:diss_enr}
where $m_q$ is mass of quark and $E_B$ is the energy eigenvalue calculated from time independent Schr\"{o}dinger equation by using Neumerovs method. We have plotted the binding energy of $J/\Psi$ at a temperature $1.7T_c$ and $\Upsilon(1S)$ at a temperature $3T_c$ as a function of the magnetic field intensity in fig.\ref{binding}. The binding energies do not change much over a span of magnetic field intensity from $1-15 m^2_{\pi}$. In other words, these quarkonium states can survive at a higher temperature if there is magnetic field present in the medium. Within the specified rage of the magnetic field intensity the dissociation temperature of $J/\Psi$ and $\Upsilon(1S )$ becomes $2.73-2.94T_c$ and $8.12-8.89T_c$ respectively. In the current experimental scenario the medium temperature does not go up to $8T_c$ and therefore we have not considered the medium temperature above $500 MeV$ for the calculation of dissociation probability.\\

\begin{figure}
\includegraphics[width=0.5\textwidth]{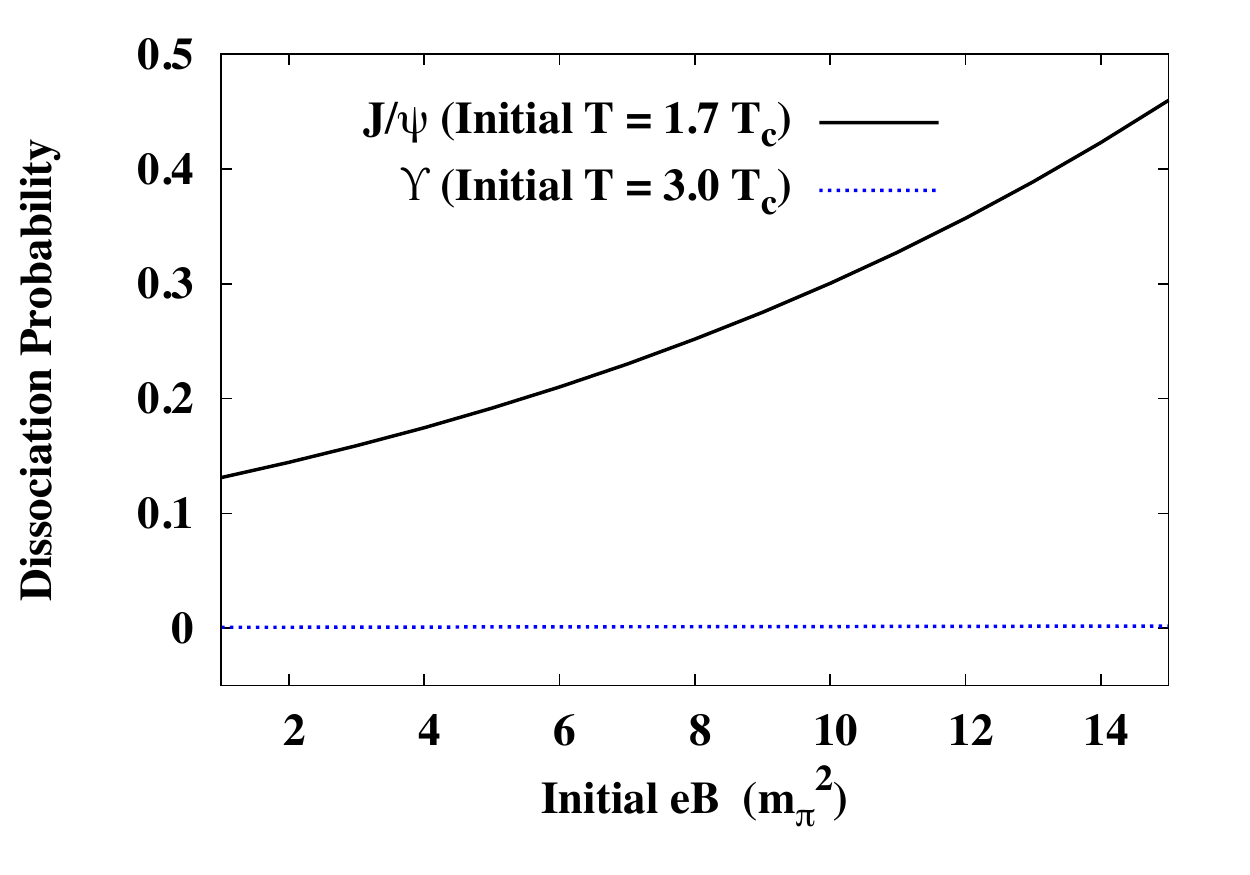}
\caption{Dissociation probability of $J/\Psi$ and $\Upsilon(1S)$ as a function of the intensity of the initial magnetic field.}
\label{dis}
\end{figure}

 We have employed first order perturbation theory to evaluate the dissociation probabilities of both the ground states first by considering a purely thermal QGP which cools off to the temperature $T_c$ of the medium and then the same has been calculated by considering the time dependent magnetic field in the evolving QGP. For $J/\Psi$, we have started at a temperature of the medium which is $1.7T_c$ and then we allow the medium temperature to reduce according to the power law given by,
 
 \begin{equation}
T(t) =T_0\big(\frac{\tau_0}{\tau_0+t} \big)^{\frac{1}{3}},
\end{equation}

with $T_0$, the initial temperature and $\tau_0$ be the equilibration time, taken to be approximately $5 fm/c$ for QGP. We have calculated the dissociation probability when the medium temperature falls off to $T_c$ from an initial valuein presence of the time dependent magnetic field. The initial value of the magnetic field is not known exactly and therefore we have used various initial values of the magnetic field intensity and have shown the dissociation probabilities as a function of initial magnetic field. The same has been done for the $\Upsilon(1S)$ state by considering the initial temperature around $3T_c$. In fig.\ref{dis} the solid black line denotes the dissociation probability of $J/\Psi$ which increases with the initial field intensity. The state $J/\Psi$ can be dissociated $12$ to $50$
percent within the range of the field intensity $1-15 m^2_{\pi}$. The dotted blue line shows that the dissociation probability for $\Upsilon(1S)$ is almost zero over the specified span of the field strength.

  Summarising the article, we conclude that due to the modification of the heavy quark potential in presence of magnetic field, the bound states $J/\Psi$ and $\Upsilon(1S)$ become more strongly bound compared to those in a pure thermal QGP. As a result, the bound states can survive much higher temperature than we have expected previously. All though $J/\Psi$ can be dissociated by making non-adiabatic transitions to the unbound states but $\Upsilon(1S)$ is still remains bound. We have estimated the dissociation probability within the limits of first order perturbation theory. For a better prediction one must solve the Schr\"{o}edinger equation for a time dependent potential which by any means seems extremely challenging.

\section{Acknowledgement}
P. B. acknowledges SERB (NPDF Scheme: PDF/2016/003837), Government of India for the financial assistance and also thanks Arpan Das, Jan-e Alam, and Surasree Mazumder for useful discussion. S.P. A. thanks Prof. T. K. Nayak for academic support.

\bibliographystyle{apsrev4-1}
\bibliography{abcd}{}

\begin{thebibliography}{25}%
\makeatletter
\providecommand \@ifxundefined [1]{%
 \@ifx{#1\undefined}
}%
\providecommand \@ifnum [1]{%
 \ifnum #1\expandafter \@firstoftwo
 \else \expandafter \@secondoftwo
 \fi
}%
\providecommand \@ifx [1]{%
 \ifx #1\expandafter \@firstoftwo
 \else \expandafter \@secondoftwo
 \fi
}%
\providecommand \natexlab [1]{#1}%
\providecommand \enquote  [1]{``#1''}%
\providecommand \bibnamefont  [1]{#1}%
\providecommand \bibfnamefont [1]{#1}%
\providecommand \citenamefont [1]{#1}%
\providecommand \href@noop [0]{\@secondoftwo}%
\providecommand \href [0]{\begingroup \@sanitize@url \@href}%
\providecommand \@href[1]{\@@startlink{#1}\@@href}%
\providecommand \@@href[1]{\endgroup#1\@@endlink}%
\providecommand \@sanitize@url [0]{\catcode `\\12\catcode `\$12\catcode
  `\&12\catcode `\#12\catcode `\^12\catcode `\_12\catcode `\%12\relax}%
\providecommand \@@startlink[1]{}%
\providecommand \@@endlink[0]{}%
\providecommand \url  [0]{\begingroup\@sanitize@url \@url }%
\providecommand \@url [1]{\endgroup\@href {#1}{\urlprefix }}%
\providecommand \urlprefix  [0]{URL }%
\providecommand \Eprint [0]{\href }%
\providecommand \doibase [0]{http://dx.doi.org/}%
\providecommand \selectlanguage [0]{\@gobble}%
\providecommand \bibinfo  [0]{\@secondoftwo}%
\providecommand \bibfield  [0]{\@secondoftwo}%
\providecommand \translation [1]{[#1]}%
\providecommand \BibitemOpen [0]{}%
\providecommand \bibitemStop [0]{}%
\providecommand \bibitemNoStop [0]{.\EOS\space}%
\providecommand \EOS [0]{\spacefactor3000\relax}%
\providecommand \BibitemShut  [1]{\csname bibitem#1\endcsname}%
\let\auto@bib@innerbib\@empty
\bibitem [{\citenamefont {Skokov}\ \emph {et~al.}(2009)\citenamefont {Skokov},
  \citenamefont {Illarionov},\ and\ \citenamefont {Toneev}}]{skokov2009}%
  \BibitemOpen
  \bibfield  {author} {\bibinfo {author} {\bibfnamefont {V.}~\bibnamefont
  {Skokov}}, \bibinfo {author} {\bibfnamefont {A.~{\relax Yu}.}\ \bibnamefont
  {Illarionov}}, \ and\ \bibinfo {author} {\bibfnamefont {V.}~\bibnamefont
  {Toneev}},\ }\href {\doibase 10.1142/S0217751X09047570} {\bibfield  {journal}
  {\bibinfo  {journal} {Int. J. Mod. Phys.}\ }\textbf {\bibinfo {volume}
  {A24}},\ \bibinfo {pages} {5925} (\bibinfo {year} {2009})},\ \Eprint
  {http://arxiv.org/abs/0907.1396} {arXiv:0907.1396 [nucl-th]} \BibitemShut
  {NoStop}%
\bibitem [{\citenamefont {Kharzeev}\ \emph {et~al.}(2008)\citenamefont
  {Kharzeev}, \citenamefont {McLerran},\ and\ \citenamefont
  {Warringa}}]{kharzeev2007}%
  \BibitemOpen
  \bibfield  {author} {\bibinfo {author} {\bibfnamefont {D.~E.}\ \bibnamefont
  {Kharzeev}}, \bibinfo {author} {\bibfnamefont {L.~D.}\ \bibnamefont
  {McLerran}}, \ and\ \bibinfo {author} {\bibfnamefont {H.~J.}\ \bibnamefont
  {Warringa}},\ }\href {\doibase 10.1016/j.nuclphysa.2008.02.298} {\bibfield
  {journal} {\bibinfo  {journal} {Nucl. Phys.}\ }\textbf {\bibinfo {volume}
  {A803}},\ \bibinfo {pages} {227} (\bibinfo {year} {2008})},\ \Eprint
  {http://arxiv.org/abs/0711.0950} {arXiv:0711.0950 [hep-ph]} \BibitemShut
  {NoStop}%
\bibitem [{\citenamefont {Mueller}\ \emph {et~al.}(2014)\citenamefont
  {Mueller}, \citenamefont {Bonnet},\ and\ \citenamefont
  {Fischer}}]{Mueller:2014tea}%
  \BibitemOpen
  \bibfield  {author} {\bibinfo {author} {\bibfnamefont {N.}~\bibnamefont
  {Mueller}}, \bibinfo {author} {\bibfnamefont {J.~A.}\ \bibnamefont {Bonnet}},
  \ and\ \bibinfo {author} {\bibfnamefont {C.~S.}\ \bibnamefont {Fischer}},\
  }\href {\doibase 10.1103/PhysRevD.89.094023} {\bibfield  {journal} {\bibinfo
  {journal} {Phys. Rev.}\ }\textbf {\bibinfo {volume} {D89}},\ \bibinfo {pages}
  {094023} (\bibinfo {year} {2014})},\ \Eprint {http://arxiv.org/abs/1401.1647}
  {arXiv:1401.1647 [hep-ph]} \BibitemShut {NoStop}%
\bibitem [{\citenamefont {Miransky}\ and\ \citenamefont
  {Shovkovy}(2015)}]{Miransky:2015ava}%
  \BibitemOpen
  \bibfield  {author} {\bibinfo {author} {\bibfnamefont {V.~A.}\ \bibnamefont
  {Miransky}}\ and\ \bibinfo {author} {\bibfnamefont {I.~A.}\ \bibnamefont
  {Shovkovy}},\ }\href {\doibase 10.1016/j.physrep.2015.02.003} {\bibfield
  {journal} {\bibinfo  {journal} {Phys. Rept.}\ }\textbf {\bibinfo {volume}
  {576}},\ \bibinfo {pages} {1} (\bibinfo {year} {2015})},\ \Eprint
  {http://arxiv.org/abs/1503.00732} {arXiv:1503.00732 [hep-ph]} \BibitemShut
  {NoStop}%
\bibitem [{\citenamefont {Kharzeev}\ and\ \citenamefont
  {Zhitnitsky}(2007)}]{Kharzeev:2007tn}%
  \BibitemOpen
  \bibfield  {author} {\bibinfo {author} {\bibfnamefont {D.}~\bibnamefont
  {Kharzeev}}\ and\ \bibinfo {author} {\bibfnamefont {A.}~\bibnamefont
  {Zhitnitsky}},\ }\href {\doibase 10.1016/j.nuclphysa.2007.10.001} {\bibfield
  {journal} {\bibinfo  {journal} {Nucl. Phys.}\ }\textbf {\bibinfo {volume}
  {A797}},\ \bibinfo {pages} {67} (\bibinfo {year} {2007})},\ \Eprint
  {http://arxiv.org/abs/0706.1026} {arXiv:0706.1026 [hep-ph]} \BibitemShut
  {NoStop}%
\bibitem [{\citenamefont {Bloczynski}\ \emph {et~al.}(2013)\citenamefont
  {Bloczynski}, \citenamefont {Huang}, \citenamefont {Zhang},\ and\
  \citenamefont {Liao}}]{bloczynski2012}%
  \BibitemOpen
  \bibfield  {author} {\bibinfo {author} {\bibfnamefont {J.}~\bibnamefont
  {Bloczynski}}, \bibinfo {author} {\bibfnamefont {X.-G.}\ \bibnamefont
  {Huang}}, \bibinfo {author} {\bibfnamefont {X.}~\bibnamefont {Zhang}}, \ and\
  \bibinfo {author} {\bibfnamefont {J.}~\bibnamefont {Liao}},\ }\href {\doibase
  10.1016/j.physletb.2012.12.030} {\bibfield  {journal} {\bibinfo  {journal}
  {Phys. Lett.}\ }\textbf {\bibinfo {volume} {B718}},\ \bibinfo {pages} {1529}
  (\bibinfo {year} {2013})},\ \Eprint {http://arxiv.org/abs/1209.6594}
  {arXiv:1209.6594 [nucl-th]} \BibitemShut {NoStop}%
\bibitem [{\citenamefont {Asakawa}\ \emph {et~al.}(2010)\citenamefont
  {Asakawa}, \citenamefont {Majumder},\ and\ \citenamefont
  {Muller}}]{Asakawa:2010bu}%
  \BibitemOpen
  \bibfield  {author} {\bibinfo {author} {\bibfnamefont {M.}~\bibnamefont
  {Asakawa}}, \bibinfo {author} {\bibfnamefont {A.}~\bibnamefont {Majumder}}, \
  and\ \bibinfo {author} {\bibfnamefont {B.}~\bibnamefont {Muller}},\ }\href
  {\doibase 10.1103/PhysRevC.81.064912} {\bibfield  {journal} {\bibinfo
  {journal} {Phys. Rev.}\ }\textbf {\bibinfo {volume} {C81}},\ \bibinfo {pages}
  {064912} (\bibinfo {year} {2010})},\ \Eprint {http://arxiv.org/abs/1003.2436}
  {arXiv:1003.2436 [hep-ph]} \BibitemShut {NoStop}%
\bibitem [{\citenamefont {Bonati}\ \emph {et~al.}(2015)\citenamefont {Bonati},
  \citenamefont {D'Elia},\ and\ \citenamefont {Rucci}}]{bonati2015}%
  \BibitemOpen
  \bibfield  {author} {\bibinfo {author} {\bibfnamefont {C.}~\bibnamefont
  {Bonati}}, \bibinfo {author} {\bibfnamefont {M.}~\bibnamefont {D'Elia}}, \
  and\ \bibinfo {author} {\bibfnamefont {A.}~\bibnamefont {Rucci}},\ }\href
  {\doibase 10.1103/PhysRevD.92.054014} {\bibfield  {journal} {\bibinfo
  {journal} {Phys. Rev.}\ }\textbf {\bibinfo {volume} {D92}},\ \bibinfo {pages}
  {054014} (\bibinfo {year} {2015})},\ \Eprint
  {http://arxiv.org/abs/1506.07890} {arXiv:1506.07890 [hep-ph]} \BibitemShut
  {NoStop}%
\bibitem [{\citenamefont {Guo}\ \emph {et~al.}(2015)\citenamefont {Guo},
  \citenamefont {Shi}, \citenamefont {Xu}, \citenamefont {Xu},\ and\
  \citenamefont {Zhuang}}]{guo2015}%
  \BibitemOpen
  \bibfield  {author} {\bibinfo {author} {\bibfnamefont {X.}~\bibnamefont
  {Guo}}, \bibinfo {author} {\bibfnamefont {S.}~\bibnamefont {Shi}}, \bibinfo
  {author} {\bibfnamefont {N.}~\bibnamefont {Xu}}, \bibinfo {author}
  {\bibfnamefont {Z.}~\bibnamefont {Xu}}, \ and\ \bibinfo {author}
  {\bibfnamefont {P.}~\bibnamefont {Zhuang}},\ }\href {\doibase
  10.1016/j.physletb.2015.10.038} {\bibfield  {journal} {\bibinfo  {journal}
  {Phys. Lett.}\ }\textbf {\bibinfo {volume} {B751}},\ \bibinfo {pages} {215}
  (\bibinfo {year} {2015})},\ \Eprint {http://arxiv.org/abs/1502.04407}
  {arXiv:1502.04407 [hep-ph]} \BibitemShut {NoStop}%
\bibitem [{\citenamefont {Yang}\ and\ \citenamefont {Muller}(2012)}]{yang2011}%
  \BibitemOpen
  \bibfield  {author} {\bibinfo {author} {\bibfnamefont {D.-L.}\ \bibnamefont
  {Yang}}\ and\ \bibinfo {author} {\bibfnamefont {B.}~\bibnamefont {Muller}},\
  }\href {\doibase 10.1088/0954-3899/39/1/015007} {\bibfield  {journal}
  {\bibinfo  {journal} {J. Phys.}\ }\textbf {\bibinfo {volume} {G39}},\
  \bibinfo {pages} {015007} (\bibinfo {year} {2012})},\ \Eprint
  {http://arxiv.org/abs/1108.2525} {arXiv:1108.2525 [hep-ph]} \BibitemShut
  {NoStop}%
\bibitem [{\citenamefont {Filip}(2013)}]{filip2013}%
  \BibitemOpen
  \bibfield  {author} {\bibinfo {author} {\bibfnamefont {P.}~\bibnamefont
  {Filip}},\ }\bibfield  {booktitle} {\emph {\bibinfo {booktitle}
  {{Proceedings, 8th International Workshop on Critical Point and Onset of
  Deconfinement (CPOD 2013): Napa, CA, USA, March 11-15, 2013}}},\ }\href@noop
  {} {\bibfield  {journal} {\bibinfo  {journal} {PoS}\ }\textbf {\bibinfo
  {volume} {CPOD2013}},\ \bibinfo {pages} {035} (\bibinfo {year}
  {2013})}\BibitemShut {NoStop}%
\bibitem [{\citenamefont {Alford}\ and\ \citenamefont
  {Strickland}(2013)}]{michael2013}%
  \BibitemOpen
  \bibfield  {author} {\bibinfo {author} {\bibfnamefont {J.}~\bibnamefont
  {Alford}}\ and\ \bibinfo {author} {\bibfnamefont {M.}~\bibnamefont
  {Strickland}},\ }\href {\doibase 10.1103/PhysRevD.88.105017} {\bibfield
  {journal} {\bibinfo  {journal} {Phys. Rev.}\ }\textbf {\bibinfo {volume}
  {D88}},\ \bibinfo {pages} {105017} (\bibinfo {year} {2013})},\ \Eprint
  {http://arxiv.org/abs/1309.3003} {arXiv:1309.3003 [hep-ph]} \BibitemShut
  {NoStop}%
\bibitem [{\citenamefont {Karshenboim}(2004)}]{karshenboim2003}%
  \BibitemOpen
  \bibfield  {author} {\bibinfo {author} {\bibfnamefont {S.~G.}\ \bibnamefont
  {Karshenboim}},\ }\bibfield  {booktitle} {\emph {\bibinfo {booktitle}
  {{Positronium physics. Proceedings, 1st International Workshop, Zuerich,
  Switzerland, May 30-31, 2003}}},\ }\href {\doibase 10.1142/S0217751X04020142}
  {\bibfield  {journal} {\bibinfo  {journal} {Int. J. Mod. Phys.}\ }\textbf
  {\bibinfo {volume} {A19}},\ \bibinfo {pages} {3879} (\bibinfo {year}
  {2004})},\ \Eprint {http://arxiv.org/abs/hep-ph/0310099}
  {arXiv:hep-ph/0310099 [hep-ph]} \BibitemShut {NoStop}%
\bibitem [{\citenamefont {Dutta}\ and\ \citenamefont
  {Mazumder}(2017)}]{Dutta:2017pya}%
  \BibitemOpen
  \bibfield  {author} {\bibinfo {author} {\bibfnamefont {N.}~\bibnamefont
  {Dutta}}\ and\ \bibinfo {author} {\bibfnamefont {S.}~\bibnamefont
  {Mazumder}},\ }\href@noop {} {\  (\bibinfo {year} {2017})},\ \Eprint
  {http://arxiv.org/abs/1704.04094} {arXiv:1704.04094 [nucl-th]} \BibitemShut
  {NoStop}%
\bibitem [{\citenamefont {Marasinghe}\ and\ \citenamefont
  {Tuchin}(2011)}]{tuchin2011}%
  \BibitemOpen
  \bibfield  {author} {\bibinfo {author} {\bibfnamefont {K.}~\bibnamefont
  {Marasinghe}}\ and\ \bibinfo {author} {\bibfnamefont {K.}~\bibnamefont
  {Tuchin}},\ }\href {\doibase 10.1103/PhysRevC.84.044908} {\bibfield
  {journal} {\bibinfo  {journal} {Phys. Rev.}\ }\textbf {\bibinfo {volume}
  {C84}},\ \bibinfo {pages} {044908} (\bibinfo {year} {2011})},\ \Eprint
  {http://arxiv.org/abs/1103.1329} {arXiv:1103.1329 [hep-ph]} \BibitemShut
  {NoStop}%
\bibitem [{\citenamefont {Das}\ \emph {et~al.}(2017)\citenamefont {Das},
  \citenamefont {Dave}, \citenamefont {Saumia},\ and\ \citenamefont
  {Srivastava}}]{ajit2017}%
  \BibitemOpen
  \bibfield  {author} {\bibinfo {author} {\bibfnamefont {A.}~\bibnamefont
  {Das}}, \bibinfo {author} {\bibfnamefont {S.~S.}\ \bibnamefont {Dave}},
  \bibinfo {author} {\bibfnamefont {P.~S.}\ \bibnamefont {Saumia}}, \ and\
  \bibinfo {author} {\bibfnamefont {A.~M.}\ \bibnamefont {Srivastava}},\
  }\href@noop {} {\  (\bibinfo {year} {2017})},\ \Eprint
  {http://arxiv.org/abs/1703.08162} {arXiv:1703.08162 [hep-ph]} \BibitemShut
  {NoStop}%
\bibitem [{\citenamefont {Dutta}\ and\ \citenamefont
  {Borghini}(2015)}]{Dutta:2012nw}%
  \BibitemOpen
  \bibfield  {author} {\bibinfo {author} {\bibfnamefont {N.}~\bibnamefont
  {Dutta}}\ and\ \bibinfo {author} {\bibfnamefont {N.}~\bibnamefont
  {Borghini}},\ }\href {\doibase 10.1142/S0217732315502053} {\bibfield
  {journal} {\bibinfo  {journal} {Mod. Phys. Lett.}\ }\textbf {\bibinfo
  {volume} {A30}},\ \bibinfo {pages} {1550205} (\bibinfo {year} {2015})},\
  \Eprint {http://arxiv.org/abs/1206.2149} {arXiv:1206.2149 [nucl-th]}
  \BibitemShut {NoStop}%
\bibitem [{\citenamefont {Bagchi}\ and\ \citenamefont
  {Srivastava}(2015)}]{Bagchi:2014jya}%
  \BibitemOpen
  \bibfield  {author} {\bibinfo {author} {\bibfnamefont {P.}~\bibnamefont
  {Bagchi}}\ and\ \bibinfo {author} {\bibfnamefont {A.~M.}\ \bibnamefont
  {Srivastava}},\ }\href {\doibase 10.1142/S021773231550162X} {\bibfield
  {journal} {\bibinfo  {journal} {Mod. Phys. Lett.}\ }\textbf {\bibinfo
  {volume} {A30}},\ \bibinfo {pages} {1550162} (\bibinfo {year} {2015})},\
  \Eprint {http://arxiv.org/abs/1411.5596} {arXiv:1411.5596 [hep-ph]}
  \BibitemShut {NoStop}%
\bibitem [{\citenamefont {Schwinger}(1951)}]{Schwinger1951}%
  \BibitemOpen
  \bibfield  {author} {\bibinfo {author} {\bibfnamefont {J.~S.}\ \bibnamefont
  {Schwinger}},\ }\href {\doibase 10.1103/PhysRev.82.664} {\bibfield  {journal}
  {\bibinfo  {journal} {Phys. Rev.}\ }\textbf {\bibinfo {volume} {82}},\
  \bibinfo {pages} {664} (\bibinfo {year} {1951})},\ \bibinfo {note}
  {[,116(1951)]}\BibitemShut {NoStop}%
\bibitem [{\citenamefont {Hasan}\ \emph {et~al.}(2017)\citenamefont {Hasan},
  \citenamefont {Chatterjee},\ and\ \citenamefont {Patra}}]{Hasan:2017fmf}%
  \BibitemOpen
  \bibfield  {author} {\bibinfo {author} {\bibfnamefont {M.}~\bibnamefont
  {Hasan}}, \bibinfo {author} {\bibfnamefont {B.}~\bibnamefont {Chatterjee}}, \
  and\ \bibinfo {author} {\bibfnamefont {B.~K.}\ \bibnamefont {Patra}},\ }\href
  {\doibase 10.1140/epjc/s10052-017-5346-z} {\bibfield  {journal} {\bibinfo
  {journal} {Eur. Phys. J.}\ }\textbf {\bibinfo {volume} {C77}},\ \bibinfo
  {pages} {767} (\bibinfo {year} {2017})},\ \Eprint
  {http://arxiv.org/abs/1703.10508} {arXiv:1703.10508 [hep-ph]} \BibitemShut
  {NoStop}%
\bibitem [{\citenamefont {Hasan}\ \emph {et~al.}(2018)\citenamefont {Hasan},
  \citenamefont {Patra}, \citenamefont {Chatterjee},\ and\ \citenamefont
  {Bagchi}}]{Hasan:2018kvx}%
  \BibitemOpen
  \bibfield  {author} {\bibinfo {author} {\bibfnamefont {M.}~\bibnamefont
  {Hasan}}, \bibinfo {author} {\bibfnamefont {B.~K.}\ \bibnamefont {Patra}},
  \bibinfo {author} {\bibfnamefont {B.}~\bibnamefont {Chatterjee}}, \ and\
  \bibinfo {author} {\bibfnamefont {P.}~\bibnamefont {Bagchi}},\ }\href@noop {}
  {\  (\bibinfo {year} {2018})},\ \Eprint {http://arxiv.org/abs/1802.06874}
  {arXiv:1802.06874 [hep-ph]} \BibitemShut {NoStop}%
\bibitem [{\citenamefont {Andreichikov}\ \emph {et~al.}(2013)\citenamefont
  {Andreichikov}, \citenamefont {Orlovsky},\ and\ \citenamefont
  {Simonov}}]{Andreichikov:2012xe}%
  \BibitemOpen
  \bibfield  {author} {\bibinfo {author} {\bibfnamefont {M.~A.}\ \bibnamefont
  {Andreichikov}}, \bibinfo {author} {\bibfnamefont {V.~D.}\ \bibnamefont
  {Orlovsky}}, \ and\ \bibinfo {author} {\bibfnamefont {{\relax Yu}.~A.}\
  \bibnamefont {Simonov}},\ }\href {\doibase 10.1103/PhysRevLett.110.162002}
  {\bibfield  {journal} {\bibinfo  {journal} {Phys. Rev. Lett.}\ }\textbf
  {\bibinfo {volume} {110}},\ \bibinfo {pages} {162002} (\bibinfo {year}
  {2013})},\ \Eprint {http://arxiv.org/abs/1211.6568} {arXiv:1211.6568
  [hep-ph]} \BibitemShut {NoStop}%
\bibitem [{\citenamefont {Ferrer}\ \emph {et~al.}(2015)\citenamefont {Ferrer},
  \citenamefont {de~la Incera},\ and\ \citenamefont {Wen}}]{Ferrer:2014qka}%
  \BibitemOpen
  \bibfield  {author} {\bibinfo {author} {\bibfnamefont {E.~J.}\ \bibnamefont
  {Ferrer}}, \bibinfo {author} {\bibfnamefont {V.}~\bibnamefont {de~la
  Incera}}, \ and\ \bibinfo {author} {\bibfnamefont {X.~J.}\ \bibnamefont
  {Wen}},\ }\href {\doibase 10.1103/PhysRevD.91.054006} {\bibfield  {journal}
  {\bibinfo  {journal} {Phys. Rev.}\ }\textbf {\bibinfo {volume} {D91}},\
  \bibinfo {pages} {054006} (\bibinfo {year} {2015})},\ \Eprint
  {http://arxiv.org/abs/1407.3503} {arXiv:1407.3503 [nucl-th]} \BibitemShut
  {NoStop}%
\bibitem [{\citenamefont {Tuchin}(2016)}]{Tuchin:2015oka}%
  \BibitemOpen
  \bibfield  {author} {\bibinfo {author} {\bibfnamefont {K.}~\bibnamefont
  {Tuchin}},\ }\href {\doibase 10.1103/PhysRevC.93.014905} {\bibfield
  {journal} {\bibinfo  {journal} {Phys. Rev.}\ }\textbf {\bibinfo {volume}
  {C93}},\ \bibinfo {pages} {014905} (\bibinfo {year} {2016})},\ \Eprint
  {http://arxiv.org/abs/1508.06925} {arXiv:1508.06925 [hep-ph]} \BibitemShut
  {NoStop}%
\bibitem [{\citenamefont {M\'ocsy}\ and\ \citenamefont
  {Petreczky}(2007)}]{PhysRevLett.99.211602}%
  \BibitemOpen
  \bibfield  {author} {\bibinfo {author} {\bibfnamefont {A.}~\bibnamefont
  {M\'ocsy}}\ and\ \bibinfo {author} {\bibfnamefont {P.}~\bibnamefont
  {Petreczky}},\ }\href {\doibase 10.1103/PhysRevLett.99.211602} {\bibfield
  {journal} {\bibinfo  {journal} {Phys. Rev. Lett.}\ }\textbf {\bibinfo
  {volume} {99}},\ \bibinfo {pages} {211602} (\bibinfo {year}
  {2007})}\BibitemShut {NoStop}%
\end{thebibliography}%

\end{document}